  \documentstyle{article}
  \topmargin - 0.5 cm 
\begin{document}

\vspace*{1.2 true cm}

% \\[1.5 true cm]

\noindent{\large{\bf ON SUPERSYMMETRIC QUANTUM MECHANICS}}

\vspace{0.8 true cm}

% \bigskip
% \bigskip

\noindent{M.R. KIBLER}

\noindent{Institut de Physique Nucl\'eaire de Lyon}

\noindent{IN2P3-CNRS et Universit\'e Claude Bernard}

% \noindent{43, Boulevard du 11 Novembre 1918}

\noindent{F-69622 Villeurbanne Cedex, France}

\noindent{and} 
 
\noindent{M. DAOUD}

\noindent{D\'epartement de Physique}

\noindent{Facult\'e des Sciences}

\noindent{Universit\'e Ibnou Zohr}

\noindent{B.P. 28/S, Agadir, Morocco}

% \bigskip
% \bigskip
% \bigskip

\vspace{1.2 true cm}

%
% \hfill
% \begin{minipage}{11.5 true cm}

\noindent{\bf Abstract.}
This paper constitutes a review on ${\cal N}=2$ 
fractional supersymmetric Quantum Mechanics of 
order $k$. The presentation is based on the 
introduction of a generalized Weyl-Heisenberg 
algebra $W_k$. It is shown how a general 
Hamiltonian can be associated with the algebra 
$W_k$. This general Hamiltonian covers 
various supersymmetrical versions 
of dynamical systems (Morse system, P\"oschl-Teller system, 
fractional supersymmetric
oscillator of order $k$, etc.). The case of ordinary supersymmetric Quantum
Mechanics corresponds to $k=2$. A connection between 
fractional supersymmetric Quantum Mechanics
and ordinary supersymmetric Quantum
Mechanics is briefly described. 
A realization of the algebra 
$W_k$, of the ${\cal N}=2$ supercharges and of the
corresponding Hamiltonian is given in terms of 
deformed-bosons and $k$-fermions as well as in terms 
of differential operators.

%
%\end{minipage}

\bigskip
\bigskip
\bigskip
\bigskip
\bigskip
\bigskip
\bigskip
\bigskip
\bigskip
\bigskip
\bigskip
\bigskip

\thispagestyle{empty}

\noindent Review paper to be published in:

\noindent {\bf Fundamental World of Quantum Chemistry} 

\noindent A Tribute to the Memory of Per-Olov L\"owdin, Volume 3

\noindent E. Brandas and E.S. Kryachko (Eds.), Springer-Verlag, Berlin, 2004.

\newpage 

% \bigskip
% \bigskip
% \bigskip

\vspace*{1.2 true cm}

% \\[1.5 true cm]

\noindent{\large{\bf ON SUPERSYMMETRIC QUANTUM MECHANICS}}

\vspace{0.8 true cm}

% \bigskip
% \bigskip

\noindent{M.R. KIBLER}

\noindent{Institut de Physique Nucl\'eaire de Lyon}

\noindent{IN2P3-CNRS et Universit\'e Claude Bernard}

% \noindent{43, Boulevard du 11 Novembre 1918}

\noindent{F-69622 Villeurbanne Cedex, France}

\noindent{and} 
 
\noindent{M. DAOUD}

\noindent{D\'epartement de Physique}

\noindent{Facult\'e des Sciences}

\noindent{Universit\'e Ibnou Zohr}

\noindent{B.P. 28/S, Agadir, Morocco}

% \bigskip
% \bigskip
% \bigskip

\vspace{1.2 true cm}

\hfill
\begin{minipage}{11.5 true cm}

\noindent{\bf Abstract.}
This paper constitutes a review on ${\cal N}=2$ 
fractional supersymmetric Quantum Mechanics of 
order $k$. The presentation is based on the 
introduction of a generalized Weyl-Heisenberg 
algebra $W_k$. It is shown how a general 
Hamiltonian can be associated with the algebra 
$W_k$. This general Hamiltonian covers 
various supersymmetrical versions 
of dynamical systems (Morse system, P\"oschl-Teller system, 
fractional supersymmetric
oscillator of order $k$, etc.). The case of ordinary supersymmetric Quantum
Mechanics corresponds to $k=2$. A connection between 
fractional supersymmetric Quantum Mechanics
and ordinary supersymmetric Quantum
Mechanics is briefly described. 
A realization of the algebra 
$W_k$, of the ${\cal N}=2$ supercharges and of the
corresponding Hamiltonian is given in terms of 
deformed-bosons and $k$-fermions as well as in terms 
of differential operators.

\end{minipage}

% \bigskip
% \bigskip
% \bigskip
% \bigskip

\section{Introducing supersymmetry}
 
Supersymmetry (SUperSYmmetry or SUSY) 
can be defined as a symmetry between bosons and fermions (as considered 
as elementary particles or simply as degrees of freedom). In other words,
SUSY is based on the postulated existence of operators $Q_{\alpha}$ which
transform a bosonic field into a fermionic field and {\em vice versa}.  
In  the  context  of  quantum  mechanics, such symmetry operators induce 
a $Z_2$-grading of the Hilbert space of quantum states. In a more 
general way, fractional SUSY corresponds to a $Z_k$-grading for which the Hilbert
space involves both bosonic degrees of freedom (associated with bosons) and
$k$-fermionic degrees of freedom (associated with $k$-fermions to be
described below) with $k \in {\bf N} \setminus \{ 0 , 1 , 2 \}$~; the case
$k=2$ corresponds to ordinary SUSY.

The concept of SUSY is very useful in Quantum Physics and 
Quantum Chemistry. It was first introduced in elementary 
particle physics on the basis of the unification of internal
and external symmetries [1,2]. More precisely, 
SUSY goes back to the sixtees when
many attemps were done in order to unify {\em external symmetries} (described by
the Poincar\'e group) and {\em internal symmetries} (described by gauge groups)
{\em for elementary particles}. These attempts 
led to a no-go theorem by Coleman and
Mandula in 1967 [1] which states that, 
 under reasonable assumptions concerning the 
$S$-matrix, the unification of internal and external symmetries can be achieved solely 
through the introduction of the direct product of the Poincar\'e group with the relevant
gauge group. In conclusion, this unification brings nothing new since it simply amounts to
consider separately the two kinds of symmetries. A way to escape this no-go theorem was
proposed by Haag, Lopuszanski and Sohnius in 1975 [2]: 
The remedy consists in replacing
the Poincar\'e group by an extended Poincar\'e group, 
the Poincar\'e {\em supergroup} (or $Z_2$-{\em graded} 
Poincar\'e {\em group}). 

Let us briefly discuss how to introduce the    Poincar\'e   supergroup. We know that the
Poincar\'e group has ten generators (six $M_{\mu \nu} \equiv - M_{\nu \mu}$ and four
$P_{\mu}$ with $\mu,\nu \in \{ 0,1,2,3 \}$): Three $M_{\mu \nu}$ describe ordinary 
rotations, three $M_{\mu \nu}$ describe special Lorentz 
transformations and the four $P_{\mu}$ describe space-time translations. The Lie algebra
of the Poincar\'e group is characterized by the commutation relations  
$$
[M_{\mu \nu} , M_{\rho \sigma}] = -{\rm i}
(g_{\mu \rho}  M_{\nu \sigma} + 
 g_{\nu \sigma}M_{\mu \rho} -
 g_{\mu \sigma}M_{\nu \rho} -
 g_{\nu \rho}  M_{\mu \sigma}) 
$$
$$
\quad [M_{\mu \nu} , P_{\lambda}] = -{\rm i}
(g_{\mu \lambda} P_{\nu} - 
 g_{\nu \lambda} P_{\mu}), 
\quad [P_{\mu} , P_{\nu}] = 0
$$
where $(g_{\mu \nu}) = {\rm diag}(1, -1, -1, -1)$. The 
minimal extension of the Poincar\'e group into a  
Poincar\'e   supergroup gives rise to a Lie {\em superalgebra} (or $Z_2$-{\em 
graded} Lie {\em algebra}) involving the ten generators $M_{\mu \nu}$  and  $P_{\mu}$ plus four
new generators $Q_{\alpha}$ (with $\alpha \in \{ 1,2,3,4 \}$) referred to as supercharges.
The Lie superalgebra of the Poincar\'e supergroup is then described
by the commutation relations above plus the additional commutation relations
$$
[M_{\mu \nu} , Q_{\alpha}] = \frac{\rm i}{4} 
\left( [\gamma_{\mu} , \gamma_{\nu}] \right)_{\alpha \beta} Q_{\beta}, 
\quad [P_{\mu} , Q_{\alpha}] = 0
$$
and the anticommutation relations
$$
\{ Q_{\alpha} , Q_{\beta} \} = -2 \left( \gamma^{\mu} C \right)_{\alpha \beta}P_{\mu}
$$
where the $\gamma$'s are Dirac matrices, $C$ the charge 
conjugation matrix and $\alpha,\beta \in \{ 1,2,3,4 \}$.

What is the consequence of this increase of symmetries 
(i.e., passing from 10 to 14 generators)
and of the introduction of anticommutators~? In general, an increase of symmetry yields an
increase of degeneracies. For instance, in condensed matter physics, passing from    
the tetragonal symmetry to the cubical symmetry leads, from a situation where the degrees
of degeneracy are 1 and 2, to a situation where the degrees of degeneracy are 1, 2 and 3
(in the absence of accidental degeneracies). Another possible consequence of the
increase of symmetry is the occurrence of new states or new particles. For instance,
the number of particles is doubled when 
going from the Schr\"odinger equation (with Galilean invariance) to the Dirac equation
(with Lorentz invariance): With each particle of mass $m$, spin $S$ and charge $e$ is
associated an antiparticle of mass $m$, spin $S$ and charge $-e$. In a similar way,
when passing from the  Poincar\'e group to the  Poincar\'e   supergroup, we associate 
with a known particle of mass $m$, spin $S$ and charge $e$ a new particle of mass $m'$,
spin $S' = \left| S \pm \frac{1}{2} \right|$ and charge $e$. 
The particle and the new particle, called a {\em sparticle} or a
{\em particlino} (according to whether as the known 
particle is a fermion or a boson), are accommodated 
in a given irreducible representation of the $Z_2$-graded Poincar\'e
group [3]. Consequently, they should have the same mass. However,
we have $m' \not= m$ because 
SUSY is a broken symmetry. In terms
of field theory, the sparticle or particlino field results from the action of a
supercharge $Q_{\alpha}$ on the particle field (and {\em vice versa}):
$$
Q_{\alpha} : {\rm fermion} \mapsto {\rm sfermion} = {\rm boson} 
$$
$$
Q_{\alpha} : {\rm boson}   \mapsto {\rm bosino}   = {\rm fermion} 
$$
(see also Refs.~3 and 4). We 
 thus speak of a selectron (a particle of spin $0$ associated with the electron) and 
              of a photino   (a particle of spin $\frac{1}{2}$ 
                                                     associated with the photon).

The experimental evidence for SUSY is not yet firmly established. Some 
arguments in favour of SUSY come from: (i) condensed matter physics with
the fractional quantum Hall effect 
and high temperature superconductivity [5];
(ii) nuclear physics where SUSY could
connect the complex structure of odd-odd nuclei to 
much simpler even-even and odd-$A$ systems [6];
and (iii) high energy physics 
 (especially in the search of supersymmetric particles and the lighter
 Higgs boson)
 where the observed signal around 115 GeV$/c^2$ for
a neutral Higgs boson is compatible with the hypotheses of 
SUSY [7].

It is not our intention to further discuss SUSY from the viewpoint of 
condensed matter physics, nuclear physics and elementary
particle physics. We shall rather focus our attention 
on supersymmetric Quantum Mechanics (sQM), 
 a supersymmetric quantum field theory in $D = 1+0$ dimension [8] that 
has received a great deal of attention in the last
 twenty years and is still in a state of development. 
In recent years, the investigation of quantum groups, 
with deformation parameters taken as roots of unity, 
has been a catalyst for the study of 
fractional sQM which is an extension 
of ordinary sQM. This extension
takes its motivation in the so-called intermediate or exotic statistics like:
(i) anyonic statistics in $D = 1 + 2$ dimensions connected to 
braid groups [5,9-11],
(ii) para-bosonic and para-fermionic 
 statistics in $D = 1 + 3$ dimensions connected to 
 permutation groups [12-15],
             and (iii) $q$-deformed statistics (see, for instance, [16,17]) 
             arising from $q$-deformed oscillator algebras [18-23]. Along this vein,
             intermediate statistics constitute a
  useful tool for the study of physical phenomena in condensed matter physics (e.g.,
  fractional quantum Hall effect and supraconductivity at high critical temperature).

Ordinary sQM needs two degrees of freedom: one bosonic
degree (described by a complex variable) and one fermionic degree (described by a
Grassmann variable). From a mathematical point of view, we then have a $Z_2$-grading
of the Hilbert space of physical states (involving 
bosonic and fermionic states). Fractional sQM of order $k$
is an extension of ordinary sQM for which the
$Z_2$-grading is replaced by a $Z_k$-grading with 
$k \in {\bf N} \setminus \{ 0,1,2 \}$.
The $Z_k$-grading corresponds to a bosonic degree of freedom (described again by a
complex variable) and a para-fermionic or 
$k$-fermionic degree of freedom (described by a generalized
Grassmann variable of order $k$). In other 
words, to pass from ordinary supersymmetry or ordinary sQM 
to fractional supersymmetry or fractional sQM of order $k$, 
we retain the bosonic variable and replace the fermionic variable by a
para-fermionic or $k$-fermionic variable (which can 
be represented by a $k \times k$ matrix). 

A possible approach to fractional sQM of order $k$ thus 
amounts to replace fermions by para-fermions 
of order $k-1$. This
yields para-supersymmetric Quantum Mechanics as first developed, 
with one boson and one para-fermion of order 2, by Rubakov and 
Spiridonov [24] and extended by various authors [25-30]. An 
alternative approach to 
fractional sQM of order $k$ consists in replacing fermions by $k$-fermions 
which are objects interpolating
between bosons (for $k \to \infty$) and fermions (for $k=2$) and which satisfy a
generalized Pauli exclusion principle 
according to which one cannot put more than $k-1$ particles 
on a given quantum state [31]. The $k$-fermions 
proved to be useful in describing 
Bose-Einstein condensation in low dimensions [16]
(see also Ref.~[17]). They take their origin 
in a pair of $q$- and ${\bar q}$-oscillator 
algebras (or $q$- and 
                      ${\bar q}$-uon algebras) with
$$
q = \frac{1}{{\bar q}} = 
    \exp \left( \frac{2 \pi {\rm i}}{k} \right),
\eqno (1)
$$
where $k \in {\bf N} \setminus \{ 0,1 \}$ [31,32]. Along this 
line, a fractional supersymmetric oscillator was derived 
in terms of boson and $k$-fermion operators in Ref.~[33]. 

Fractional sQM can be developed also without an explicit
introduction of $k$-fermionic degrees of freedom [34,35]. In
this respect, fractional sQM of order $k=3$ 
was worked out [35] owing to  
the introduction of a $C_{\lambda}$-extended
oscillator algebra in the framework of an extension 
of the construction of sQM with one 
bosonic degree of freedom only [34].  

The connection between fractional sQM (and thus ordinary sQM) and quantum groups 
has been worked out by several authors [36-44]
mainly with applications to exotic statistics in mind. In
 particular, LeClair and Vafa [36] studied 
the isomorphism between the affine quantum algebra 
$U_q(sl_2)$ and ${\cal N} = 2$ fractional sQM in $D = 1+1$ dimensions when $q^2$
goes to a root of unity (${\cal N}$ is the number of 
supercharges); in the special case where $q^2 \to -1$,
they recovered ordinary sQM. 

This paper is a review on fractional sQM. Ordinary sQM 
and fractional sQM can be useful in Quantum Chemistry 
from two points of view. Firstly, 
SUSY allows to deal with dynamical 
systems exhibiting fermionic (possibly $k$-fermionic) and bosonic
degrees of freedom. Secondly, it makes possible to treat in a unified 
way nonrelativistic systems controled by potentials connected via 
transformations
similar to the ones used in the factorization method. It is thus hoped that the
the present work will open an avenue of future
investigations in the applications of SUSY to Quantum Chemistry. 

The content of this paper is as follows. 
Fractional sQM of order $k$ is approached from a generalized 
Weyl-Heisenberg algebra $W_k$ (defined in Section~2). In
Section~3, a general fractional supersymmetric Hamiltonian is derived 
from the generators of $W_k$. This Hamiltonian is specialized (in Section 4) to 
the case of a fractional supersymmetric oscillator. Finally, 
differential realizations, involving
bosonic and generalized Grassmannian variables, of fractional sQM 
are given in Section~5 for some particular cases of $W_k$. Some 
concluding remarks (in Section 6) close this paper. Two appendices 
complete this paper:  
The quantum algebra $U_q(sl_2)$ with $q^k = 1$ is
connected to $W_k$ in Appendix A and a boson $+$ 
$k$-fermion decomposition of a $Q$-uon for $Q \to 
q$ is derived in Appendix B.

Throughout the present work, we use the notation $[A , B]_Q = AB - QBA$
for any complex number $Q$ and any pair of operators $A$ and $B$. 
As particular cases, we have 
$[ A , B ] \equiv
 [ A , B ]_- = [A , B]_1$ and 
$\{A , B\} \equiv
 [ A , B ]_+ = [A , B]_{-1}$ for the commutator and the anticommutator, 
respectively, of $A$ and $B$. As usual, ${\bar z}$ denotes the complex 
conjugate of the number $z$ and $A^{\dagger}$ stands for the Hermitean 
conjugate of the operator $A$.

\section{A generalized Weyl-Heisenberg algebra $W_k$}
\subsection{The algebra $W_k$}
For fixed $k$, with $k \in {\bf N} \setminus \{ 0,1 \}$, we define a  
generalized Weyl-Heisenberg algebra, denoted as $W_k$, as an algebra 
spanned by four linear operators $X_-$ (annihilation operator), 
$X_+$ (creation operator), $N$ (number operator) and 
$K$ ($Z_k$-grading operator) acting on some separable 
Hilbert space and satisfying the following relations:
$$
 [X_- , X_+] = \sum_{s=0}^{k-1} f_s(N) \> \Pi_s, 
 \eqno (2{\rm a})
$$
$$
 [N , X_-]   = - X_-,  \quad
 [N , X_+]   = + X_+,
 \eqno (2{\rm b})
$$
$$
 [K , X_+]_q = [K , X_-]_{\bar q} = 0,
 \eqno (2{\rm c})
$$
$$
 [K , N] = 0, 
 \eqno (2{\rm d})
$$
$$
 K^k = 1,
  \eqno (2{\rm e})
% \eqno (2{\rm e}((2a)-(2e)=(2PLA)))
$$
where $q$ is the $k$-th root of unity given by (1).
In Eq.~(2a), 
the $f_s$ are arbitrary functions (see below) and the
operators $\Pi_{s}$ are polynomials in $K$ defined by  
$$
\Pi_{s} = \frac{1}{k}  \>  \sum_{t=0}^{k-1}  \>  q^{-st} \> K^t
\eqno (3)
$$
for $s = 0, 1, \cdots, k-1$. Furthermore, we suppose that the operator 
$K$ is unitary ($K^{\dagger} = K^{-1}$), the operator $N$ is self-adjoint
($N^{\dagger} = N$), and the operators $X_-$ and $X_+$ are connected via
Hermitean conjugation ($X_-^{\dagger} = X_+$).  The functions 
$f_s : N \mapsto f_s(N)$ must satisfy the
constrain relation
$$
f_s(N)^{\dagger} = f_s(N)
$$
(with $s = 0, 1, \cdots, k-1$) in order that $X_+ = X_-^{\dagger}$ be verified. 

\subsection{Projection operators for $W_k$}
It is easy to show that we have the 
resolution of the identity operator
$$
\sum_{s=0}^{k-1} \Pi_s = 1
$$
and the idempotency relation
$$
\Pi_s \Pi_t = \delta (s,t) \Pi_s
$$
where $\delta$ is the Kronecker symbol. Consequently, 
the $k$ Hermitean operators $\Pi_s$ are 
projection operators for the cyclic group 
$Z_k = \{ 1, K, \cdots, K^{k-1} \}$ of order $k$ spanned by the generator $K$. In 
addition, these projection operators satisfy 
$$
\Pi_s X_+ = X_+ \Pi_{s - 1}  \Leftrightarrow X_- \Pi_s = \Pi_{s - 1} X_-
\eqno (4)
$$
with the convention $\Pi_{-1} \equiv \Pi_{k-1}$ and 
                    $\Pi_k    \equiv \Pi_0$ 
(more generally, $\Pi_{s+kn}  \equiv \Pi_{s}$ for $n \in {\bf Z}$). Note that
Eq.~(3) can be reversed in the form
$$
K^t = \sum_{s=0}^{k-1} q^{ts} \> \Pi_s
$$
with $t = 0, 1, \cdots, k-1$.

The projection operators $\Pi_s$ can also be discussed in a form 
closer to the one generally used in Quantum Chemistry. The eigenvalues of 
$K$ are $q^t$ with $t = 0, 1, \cdots, k-1$. One could easily establish
that the eigenvector $\phi_t$ associated with $q^t$ satisfies
$$
\Pi_s \phi_t = \delta (s,t) \phi_t,
$$ 
a relation that makes clearer the definition of the projection operators
$\Pi_s$.

\subsection{Representation of $W_k$}
We now consider an Hilbertean representation of the algebra $W_k$. Let
${\cal F}$ be the Hilbert-Fock space 
on which the generators of $W_k$ act. 
Since $K$ obeys the cyclicity condition $K^k = 1$, the 
operator $K$ admits the set $\{ 1, q, \cdots, q^{k-1} \}$ of eigenvalues. 
It thus makes it possible to endow the 
representation space ${\cal F}$ of the 
algebra $W_k$ with a $Z_k$-grading as
$$
 {\cal F} = \bigoplus_{s=0}^{k-1} {\cal F}_s 
 \eqno (5{\rm a})
$$  
where the subspace 
$$
{\cal F}_s = \{ | n , s \rangle : n = 1, 2, \cdots, d \},
 \eqno (5{\rm b})
$$
with
$$
K | n , s \rangle = q^{s} | n , s \rangle,
$$
is a $d$-dimensional space ($d$ can be finite or infinite). 
Therefore, to each eigenvalue $q^{s}$ (with $s = 0, 1, \cdots, k-1$)
we associate a subspace ${\cal F}_s$ of ${\cal F}$. It is evident that
$$
\Pi_s | n , t \rangle = \delta (s,t) \> | n , s \rangle
$$
and, thus, the application $\Pi_s : {\cal F} \to {\cal F}_s$ yields a
projection of ${\cal F}$ onto its subspace ${\cal F}_s$.

The action of $X_{\pm}$ and $N$ on ${\cal F}$ can be taken to be 
$$
N | n , s \rangle = 
n | n , s \rangle
$$
and
$$
X_- | n , s \rangle = \sqrt {F_s (n)} \>
                       | n-1 , s-1 \rangle, \quad s \not=0,
\eqno (6{\rm a})
$$
$$
X_- | n , 0 \rangle = \sqrt {F_s (n)} \>
                       | n-1 , k-1 \rangle, \quad s     =0,
\eqno (6{\rm b})
$$
$$
X_+ | n , s \rangle = \sqrt {F_{s+1} (n + 1)} \>
                       | n+1 , s+1 \rangle, \quad s \not=k-1,
\eqno (6{\rm c})
$$
$$
X_+ | n ,k-1\rangle = \sqrt {F_{s+1} (n + 1)} \>
                       | n+1 , 0   \rangle, \quad s     =k-1.
\eqno (6{\rm d})
$$
The function $F_s$ is a structure function that
fulfills 
$$
F_{s+1}(n + 1) - F_s(n) = f_s(n)
$$
with the initial condition $F_s(0) = 0$ 
for $s = 0, 1, \cdots, k-1$ (cf. Refs.~[45,46]). 
% Furthermore, it satisfies
% which admits the classical solution $F(N) = N$ 
% for $f_s=1$ ($s= 0, 1, \cdots, k-1$).

\subsection{A deformed-boson $+$ $k$-fermion realization of $W_k$}
\subsubsection{The realization of $W_k$}
In Section 2.4, 
the main tools consist of $k$ pairs ($b(s)_-,b(s)_+$) with 
$s = 0, 1, \cdots, k-1$ of deformed-bosons and one pair ($f_-,f_+$) 
of $k$-fermions. The operators $f_{\pm}$ satisfy (see Appendix B)
$$
 [ f_- , f_+ ]_q = 1, \quad
 f_-^k = 
 f_+^k = 0,
$$
and the operators $b(s)_{\pm}$ the commutation relation
$$
 [ b(s)_- , b(s)_+ ] = f_s(N),
 \eqno (7)
$$
where the functions $f_s$ with $s = 0, 1, \cdots, k-1$ and 
the operator $N$ occur in Eq.~(2). In addition, the pairs 
($f_-,f_+$) and ($b(s)_-,b(s)_+$) are two pairs of 
commuting operators and the operators $b(s)_{\pm}$ commute with the projection 
operators $\Pi_t$ with $s,t = 0, 1, \cdots, k-1$. Of course, we have 
$b(s)_+  = b(s)_-^{\dagger}$ but 
$f_+ \not=    f_-^{\dagger}$ except for $k = 2$. The 
$k$-fermions introduced in [31] and recently 
discussed in [32] are objects interpolating
between fermions and bosons (the case $k=2$ corresponds 
to ordinary fermions and the case $k \to \infty$ to ordinary
bosons); the $k$-fermions also share some features of the 
anyons introduced in [9-11]. We now introduce the 
linear combinations
$$
b_- = \sum_{s=0}^{k-1} b(s)_- \> \Pi_s, \quad 
b_+ = \sum_{s=0}^{k-1} b(s)_+ \> \Pi_s.
$$
It is immediate to verify that we have the commutation relation
$$
[ b_- , b_+ ] = \sum_{s=0}^{k-1} f_s(N) \> \Pi_s,
 \eqno (8)
$$
a companion of Eq.~(7). Indeed, in the case where 
$f_s=1$ ($s = 0, 1, \cdots, k-1$), the two pairs ($b_-,b_+$) 
and ($f_-,f_+$) may be considered as originating from a pair 
($a_-,a_+$) of $Q$-uons 
through the $Q$-uon $\to$ boson $+$ $k$-fermion decomposition
described in Appendix B.

   We are now in a situation to find a realization of the generators $X_-$,
$X_+$ and $K$ of the algebra $W_k$ in terms of the $b$'s and $f$'s. Let us define 
the shift operators $X_-$ and $X_+$ by (see [33])
$$
X_- = b_- \left( f_-   +    \frac{ f_+ ^{k-1}}{[[k - 1]]_q !} \right),
\eqno (9)
$$
$$
X_+ = b_+ \left( f_-   +    \frac{ f_+ ^{k-1}}{[[k - 1]]_q !} \right)^{k-1},
\eqno (10)
$$
where the $q$-deformed factorial is given by
$$
\lbrack\lbrack n \rbrack\rbrack_q ! = 
\lbrack\lbrack 1 \rbrack\rbrack_q \>
\lbrack\lbrack 2 \rbrack\rbrack_q \> \cdots \> 
\lbrack\lbrack n \rbrack\rbrack_q 
$$
for  $n \in {\bf N}^*$ (and   
$\lbrack\lbrack 0 \rbrack\rbrack_q ! = 1$) and
where the symbol $[[ \ ]]_q$ is defined by
$$
[[X]]_q = {1 - q^X \over 1 - q} 
$$
with $X$ an arbitrary operator or number.
It is also always possible to find a representation 
for which the relation $X_-^{\dagger} = X_+$ holds 
(see Section 2.4.2). 
Furthermore, we define the grading operator $K$ by
$$
K = [ f_- , f_+ ].
\eqno (11)
$$
In view of the remarkable property
$$
\left( f_- + \frac{f_+^{k-1}}{[[k-1]]_q!} \right)^k = 1,
$$ 
we obtain 
$$
[X_- , X_+] = 
[b_- , b_+].
\eqno (12)
$$
Equations (8) and (12) show that Eq.~(2a) is satisfied. 
It can be checked also 
that the operators $X_-$, $X_+$ and $K$ satisfy Eqs.~(2c) and (2e). Of course, 
Eqs.~(2b) and (2d) have to be considered as postulates. However, note that the 
operator $N$ is formally given in terms of the $b$'s by Eq.~(7). 
% $$
% F(N+1) = b_- b_+ = \sum_{s=0}^{k-1} b(s)_- b(s)_+ \> \Pi_{s},
% $$
% $$
% F(N)   = b_+ b_- = \sum_{s=0}^{k-1} b(s)_+ b(s)_- \> \Pi_{s},
% $$
% with the help of the structure function $F$ introduced in Section~2.3. 
We thus have a 
realization of the generalized Weyl-Heisenberg algebra $W_k$ by multilinear forms 
involving $k$ pairs ($b(s)_- , b(s)_+$) of deformed-boson 
operators ($s = 0, 1, \cdots, k-1$) and one pair ($f_- , f_+$) 
of $k$-fermion operators.

\subsubsection{Actions on the space ${\cal F}$} 
Equation (7) is satisfied by
$$
b(s)_- b(s)_+ = F_{s+1}(N + 1), \quad
b(s)_+ b(s)_- = F_{s  }(N    ),
$$
where the structure functions $F_s$ are connected 
to the structure constants $f_s$ via
$$
F_{s+1}(N + 1) - F_s(N) = f_s(N). 
$$
% and to the the structure function $F$ via
% $$
% F(N) = \sum_{s=0}^{k-1} F_s(N) \> \Pi_s
% $$
% (cf.~Section~2.3).

Let us consider the operators $X_-$ and $X_+$ defined by Eqs.~(9) 
and (10) 
and acting on the Hilbert-Fock space ${\cal F}$ (see Eq.~(5)). We choose the action 
of the constituent operators 
$b_{\pm}$ and $f_{\pm}$ on the state $| n , s \rangle$
to be given by
$$
b_-    | n , s \rangle = 
b(s)_- | n , s \rangle = 
\sqrt {F_s(n + \sigma - \frac{1}{2})} \> 
| n - 1 , s \rangle,
$$
$$
b_+    | n , s \rangle = 
b(s)_+ | n , s \rangle = 
\sqrt {F_s(n + \sigma + \frac{1}{2})} \> 
| n + 1 , s \rangle,
$$   
and
$$
f_- | n , s \rangle = | n , s-1 \rangle, \quad
f_- | n , 0 \rangle = 0,
$$
$$
f_+ | n , s \rangle = [[s + 1]]_q \> | n , s+1 \rangle, \quad
f_+ | n , k - 1 \rangle = 0,
$$
where $\sigma = \frac{1}{2}$, 
$n \in {\bf N}$ and 
$s= 0, 1, \cdots, k-1$. The action of 
$b_{\pm}$ is standard and the action of $f_{\pm}$ corresponds 
to $\alpha = 0$ and $\beta = 1$ in Appendix B.
Then, we can show that the relationships (6) are satisfied. In 
this representation, it is easy to prove that the
Hermitean conjugation relation $X_-^{\dagger} = X_+$ 
is true.

\subsection{Particular cases for $W_k$}
The algebra $W_k$ covers a great number of situations
encountered in the 
literature (see Refs.~[33-35,47,48]). These situations differ 
by the form given to the
right-hand side of (2a) and can be classified as follows.

(i) As a particular case, the algebra $W_2$ for $k=2$ with 
$$
[ X_- , X_+] = 1 + c \> K, \quad 
 [N , X_{\pm}]   = {\pm} X_{\pm},
$$
$$
 [ K , X_{\pm} ]_+ = 0, \quad 
 [K , N] = 0, \quad K^2 = 1,
$$
where $c$ is a real constant ($f_0 = 1 + c$, 
                              $f_1 = 1 - c$),
corresponds to the Calogero-Vasiliev [47] algebra
                considered in Ref.~[48] 
for describing a system of two anyons, 
with  an  Sl($2 , {\bf R}$) dynamical symmetry,  
subjected to an intense magnetic field
and in Ref.~[34] 
for constructing sQM without fermions. Of 
course, for $k=2$ and $c=0$ we recover 
the algebra describing the ordinary or
$Z_2$-graded supersymmetric oscillator.

If we define
$$
c_s = \frac{1}{k} \sum_{t = 0}^{k-1} q^{- ts} \> f_t(N),
\eqno (13)
$$
with the functions $f_t$ chosen in such a way that $c_s$ is independent 
of $N$ (for $s = 0, 1, \cdots, k-1$), the algebra $W_k$ defined
by
$$
[ X_- , X_+] = \sum_{s = 0}^{k-1} c_s \> K^s,
\eqno (14)
$$
together with Eqs.~(2b)-(2e), corresponds to the $C_{\lambda}$-extented 
harmonic oscillator algebra introduced in Ref.~[35] for formulating 
fractional sQM of order $k=3$.

(ii) Going back to the general case where 
$k \in {\bf N} \setminus \{ 0 , 1 \}$,
if we assume in Eq.~(2a) that $f_s = G$ is independent of $s$ 
with $G(N)^{\dagger} = G(N)$, we get
$$
[ X_- , X_+] = G(N).
\eqno (15)
$$
We refer 
the algebra $W_k$ defined by Eq.~(15) together with Eqs.~(2b)-(2e)
to as a nonlinear Weyl-Heisenberg algebra (see also Ref.~[15]). The 
latter algebra was considered by the authors as a generalization 
of the $Z_k$-graded Weyl-Heisenberg algebra describing a generalized 
fractional supersymmetric oscillator [33].

(iii) As a particular case, for $G = 1$ we have
$$
[ X_- , X_+] = 1
\eqno (16)
$$
and here we can take
$$
N = X_+ X_-.
\eqno (17)
$$
The algebra $W_k$ defined by Eqs.~(16) and (17) together with Eqs.~(2b)-(2e)
corresponds to the $Z_k$-graded 
    Weyl-Heisenberg algebra connected to the fractional
    supersymmetric oscillator studied in Ref.~[33]. 

(iv) Finally, it is to be noted that the affine quantum algebra $U_q(sl_2)$
with $q^k = 1$ can be considered as a special case of 
the generalized Weyl-Heisenberg algebra $W_k$ (see Appendix A).
This result is valid for all the representations (studied in Ref.~[49]) 
of the algebra $U_q(sl_2)$.

\section{A general supersymmetric Hamiltonian}
\subsection{Axiomatic of supersymmetry}
The axiomatic of {\em ordinary} sQM is known since 
more than 20 years. A doublet of linear operators 
$(H,Q)$, where $H$ is a self-adjoint operator 
and where 
the operators $H$ and $Q$ act on a separable 
Hilbert space and satisfy the relations 
$$
Q_- = Q,                             \quad 
Q_+ = Q^{\dagger}                    \quad 
(\Rightarrow \ Q_-^{\dagger} = Q_+), \quad 
Q_{\pm}^2 = 0 
% \eqno (1{\rm a}PLA???)
$$
$$
Q_- Q_+  +  
Q_+ Q_-  =       
H
% \eqno (1{\rm b}PLA???)
$$
$$
[H , Q_{\pm}] = 0
% \eqno (1{\rm c}PLA???)
$$
is said to define a supersymmetric quantum-mechanical 
system (see Ref.~[8]). The operator $H$ is 
referred to as 
the Hamiltonian of the system spanned by the 
supersymmetry operator $Q$. The latter operator
yields the two nilpotent operators, of order $k = 2$, 
$Q_-$ and $Q_+$. These dependent operators are 
called supercharge operators. The 
system described by the doublet $(H,Q)$ is called 
an ordinary supersymmetric quantum-mechanical
system~; it corresponds to a $Z_2$-grading
with fermionic and bosonic states.  

The preceding definition of ordinary sQM can be extended 
to {\em fractional} sQM of order $k$, with $k \in {\bf N} \setminus \{0,1,2\}$. 
Following Refs.~[24]-[28], a doublet of linear operators 
$(H,Q)_k$, with $H$ a self-adjoint operator and $Q$ a supersymmetry operator, 
acting on a separable Hilbert space and satisfying the relations 
$$
Q_- = Q,                             \quad 
Q_+ = Q^{\dagger}                    \quad 
(\Rightarrow \ Q_-^{\dagger} = Q_+), \quad 
Q_{\pm}^k = 0 
\eqno (18{\rm a})
$$
$$
Q_- ^{k-1} Q_+  +  Q_- ^{k-2} Q_+ Q_- 
                                      +   \cdots 
                                      +   Q_+ Q_- ^{k-1}
                                      =       Q_- ^{k-2} H
\eqno (18{\rm b})
$$
$$
[H , Q_{\pm}] = 0
\eqno (18{\rm c})
$$
is said to 
define a $k$-fractional supersymmetric 
quantum-mechanical system. 
The operator $H$ is the Hamiltonian of the system 
spanned by the two (dependent) supercharge operators 
$Q_-$ and $Q_+$ that are nilpotent operators of order $k$. 
This system corresponds to a $Z_k$-grading
with $k$-fermionic and bosonic states. It is clear that 
the special case $k=2$ corresponds to 
an ordinary supersymmetric quantum-mechanical
system. Note that the definition (18) corresponds to a 
${\cal N} = 2$ formulation of fractional sQM of order $k$ 
($\frac{1}{2}{\cal N}$ is the number of independent supercharges).

\subsection{Supercharges}
It is  possible  to associate a supersymmetry operator
$Q$ with a generalized Weyl-Heisenberg algebra  $W_k$.
We define the supercharge operators $Q_-$ and $Q_+$ by 
$$
% Q           \equiv 
Q_- = X_- (1 - \Pi_{1}) = (1 - \Pi_{0}) X_-, 
\eqno (19{\rm a})
$$
$$
% Q^{\dagger} \equiv 
Q_+ = X_+ (1 - \Pi_{0}) = (1 - \Pi_{1}) X_+,
\eqno (19{\rm b})
$$
or alternatively 
$$
Q_- = X_- ( \Pi_{2} + \cdots + \Pi_{k-2} + \Pi_{k-1} + \Pi_{0} ), 
\eqno (19{\rm c})
$$
$$
Q_+ = X_+ ( \Pi_{1} + \Pi_{2} + \cdots + \Pi_{k-2} + \Pi_{k-1} ). 
\eqno (19{\rm d})
$$
Indeed, we have here one of $k$, with $k \in {\bf N} \setminus \{ 0 , 1 \}$, 
possible equivalent definitions of the
supercharges $Q_-$ and $Q_+$ corresponding to the $k$ circular permutations of
the indices $0, 1, \cdots, k-1$. Obviously, we have the Hermitean conjugation
relation
$$
Q_-^{\dagger} = Q_+.
$$
By making use of the commutation relations between the projection
operators $\Pi_s$ and the shift operators $X_-$ and $X_+$ [see Eq.~(4)], 
we easily get
$$
Q_-^m = X_-^m ( \Pi_{0}   + 
                \Pi_{m+1} + 
                \Pi_{m+2} + 
                \cdots    + 
                \Pi_{k-1} )
\eqno (20{\rm a})
$$
$$
Q_+^m = X_+^m ( \Pi_{1}         + 
                \Pi_{2}         + 
                \cdots          + 
                \Pi_{k - m - 1} + 
                \Pi_{k - m} )
\eqno (20{\rm b})                
$$
for $m = 0, 1, \cdots, k-1$. By combining Eqs.~(19) and (20), we obtain
$$
Q_-^k = Q_+^k = 0. 
$$
Hence, the supercharge operators $Q_-$ and $Q_+$ are nilpotent operators of order
$k$. 

We continue with some relations of central importance for 
the derivation of a
supersymmetric Hamiltonian. The basic relations are 
$$
Q_+ Q_-^m = X_+ X_-^m ( 1 - \Pi_m ) ( \Pi_{0} + \Pi_{m+1} + \cdots + \Pi_{k-1} ) 
\eqno (21{\rm a})
$$
$$
Q_-^m Q_+ = X_-^m X_+ ( 1 - \Pi_0 ) ( \Pi_{m} + \Pi_{m+1} + \cdots + \Pi_{k-1} ) 
\eqno (21{\rm b})
$$
with $m = 0, 1, \cdots, k-1$. From Eq.~(21), we can derive the following
identities giving $Q_-^m Q_+ Q_-^{\ell}$ with $m + \ell = k - 1$.

(i) We have
$$
Q_+ Q_-^{k-1} = X_+ X_-^{k-1} \Pi_{0}
\eqno (22{\rm a})
$$
$$
Q_-^{k-1} Q_+ = X_-^{k-1} X_+ \Pi_{k-1}
\eqno (22{\rm b})
$$
in the limiting cases corresponding to $(m = 0,   \ell = k - 1)$ and 
                                       $(m = k-1, \ell = 0    )$.

(ii) Furthermore, we have 
$$
Q_-^m Q_+ Q_-^{\ell} = X_-^m X_+ X_-^{\ell} ( \Pi_{0} + \Pi_{k-1} )
\eqno (22{\rm c})
$$
with the conditions $(m \not= 0,   \ell \not= k - 1)$ and 
                    $(m \not= k-1, \ell \not= 0    )$.

\subsection{The general Hamiltonian}                                       
We are now in a position to associate a $k$-fractional 
supersymmetric quantum-mechanical system with the algebra
$W_k$ characterized by a given set of functions 
$\{ f_s : s = 0, 1, \cdots, k-1 \}$. By using Eqs.~(2), 
(19) and (22), 
we find that the most general expression of $H$ 
defined by Eq.~(18) is [50]
   $$
   H = (k-1) X_+ X_-                                                       
   - \sum_{s = 3  }^{k  } 
     \sum_{t =   2}^{s-1} (t - 1) \> f_{t}(N - s + t )         \Pi_{s}
   $$ 
   $$   
   - \sum_{s =   1}^{k-1} 
     \sum_{t =   s}^{k-1} (t - k) \> f_{t}(N - s + t )         \Pi_{s}
     \eqno (23)
   $$ 
in terms of the product $X_+ X_-$, the operators $\Pi_s$ and the functions
$f_s$. In 
the general case, we can check that 
$$
H^{\dagger} = H
\eqno (24)
$$  
and
$$
[H , Q_-] = [H , Q_+] = 0.
\eqno (25)
$$
Equations (24) and (25) show 
that the two supercharge operators $Q_-$ and $Q_+$ are two (non independent) constants 
of the motion for the
Hamiltonian system described by the self-adjoint operator $H$. 
As a result, the doublet $(H, Q)_k$ associated to $W_k$
satisfies Eq.~(18) and thus defines a $k$-fractional 
supersymmetric quantum-mechanical system.
From Eqs.~(23)-(25), it can be seen that 
the Hamiltonian $H$ is a linear combination of the
projection operators $\Pi_{s}$ with coefficients corresponding to isospectral 
Hamiltonians (or supersymmetric partners) associated with the various subspaces 
${\cal F}_s$ with $s=0, 1, \cdots, k-1$ (see Section 3.5).

The Hamiltonian $H$ 
and the supercharges $Q_-$ and $Q_+$
can be expressed by means of the 
deformed-bosons and $k$-fermions. By using the identity
$$
\Pi_s \left( f_-   +    \frac{ f_+ ^{k-1}}{[[k - 1]]_q !} \right)^{n} =
      \left( f_-   +    \frac{ f_+ ^{k-1}}{[[k - 1]]_q !} \right)^{n}
\Pi_{s+n},
$$
with $s = 0, 1, \cdots, k-1$ and $n \in {\bf N}$, the 
supercharges  $Q_-$  and  $Q_+$  can be rewritten as
$$
Q_- = \left( f_-   +    \frac{ f_+ ^{k-1}}{[[k - 1]]_q !} \right) 
\> \sum_{s=1}^{k-1} b(s)_-   \> \Pi_{s+1},
$$
$$
Q_+ = \left( f_-   +    \frac{ f_+ ^{k-1}}{[[k - 1]]_q !} \right)^{k-1} 
\> \sum_{s=1}^{k-1} b(s+1)_+ \> \Pi_s,
$$
with the convention $b(k)_+ = b(0)_+$. Then, the supersymmetric 
Hamiltonian $H$ given by Eq.~(23) assumes a form 
% the form (voir la somme sur $s$???)
% $$
%  H = (k-1)
%        \sum_{s =   0}^{k-1} F_s(N)                               \Pi_{s}   
%      - \sum_{s = 3???}^{k  } 
%        \sum_{t =   2}^{s-1} (t - 1) \> f_{t}(N - s + t )         \Pi_{s}
% $$
% $$
%      - \sum_{s =   1}^{k-1} 
%        \sum_{t =   s}^{k-1} (t - k) \> f_{t}(N - s + t )         \Pi_{s},
% $$ 
involving the operators $b(s)_{\pm}$, 
          the projection operators
$\Pi_s$ (that may be written with $k$-fermion operators), 
        and the structure constants $f_s$ with $s = 0, 1, \cdots, k-1$.

\subsection{Particular cases for the Hamiltonian}
The extended Weyl-Heisenberg algebra $W_k$ 
covers numerous algebras (see Section 2.5).
Therefore, the general expression (23) for 
the Hamiltonian  $H$ associated with $W_k$ 
can be particularized to  some interesting 
cases describing exactly solvable one-dimensional 
systems. Indeed, the particular system corresponding to a given set 
$\{ f_s : s = 0, 1, \cdots, k-1 \}$ yields, in a Schr\"odinger
picture, a particular dynamical system with a specific potential. 

(i) In the particular case $k=2$, by taking $f_0 = 1 + c$ and $f_1 = 1 - c$, 
where $c$ is a real constant, the Hamiltonian (23) gives back the one derived 
in Ref.~[34].   

More generally, by restricting the functions $f_t$ in Eq.~(23) 
to constants (independent of $N$) defined by
$$
f_s = \sum_{t=0}^{k-1} q^{st} \> c_t
$$
in terms of the constants 
$c_t$ (cf.~Eq.~(13)), the so-obtained Hamiltonian $H$
corresponds to the $C_{\lambda}$-oscillator 
fully investigated for $k=3$ in Ref.~[35]. The case
$$
\forall s \in \{ 0, 1, \cdots, k-1 \} \ : \ 
f_s(N) = f_s \mbox{ independent of } N
$$
corresponds to systems with cyclic shape-invariant potentials 
(in the sense of Ref.~[51]).

(ii) In the case $f_s = G$ (independent of $s = 0, 1, \cdots, k-1$), i.e., for
a generalized Weyl-Heisenberg algebra $W_k$ defined by (2b)-(2e) and (15), the
Hamiltonian $H$ can be written as 
$$
H   = (k-1) X_+ X_-                                                             
    - \sum_{s=2}^{k  -1}
        \sum_{t=1}^{s  -1}         G(N-t) (1 - \Pi_1 - \Pi_2 - \cdots - \Pi_{s})
$$
$$    
    + \sum_{s=1}^{k  -1}
        \sum_{t=0}^{k-s-1} (k-s-t) G(N+t) \> \Pi_{s}. 
        \eqno (26)
$$
The latter expression was derived in Ref.~[33].

(iii) The case
$$
G(N) = a N + b  \mbox{ where } (a,b) \in {\bf R}^2
$$
corresponds to systems with translational shape-invariant 
potentials (in the sense of Ref.~[52]). For instance, 
the case $(a = 0, b > 0)$ 
corresponds to the harmonic oscillator potential, 
the case $(a < 0, b > 0)$ to the Morse potential and
the case $(a > 0, b > 0)$ to the P\"oschl-Teller potential. For these 
various potentials, the part of $W_k$ spanned by $X_-$, $X_+$ and $N$ 
can be identified with the ordinary Weyl-Heisenberg algebra 
for $(a = 0, b \not= 0)$,
                  with the su(2)   Lie algebra 
for $(a < 0, b > 0)$
		  and 
		  with the su(1,1) Lie algebra 
for $(a > 0, b > 0)$. 

(iv) If $G=1$, i.e., for a Weyl-Heisenberg algebra defined by (2b)-(2e) and
(16), Eq.~(26) leads to the Hamiltonian 
$$
H = (k-1) X_+ X_-  
  + (k-1) \sum_{s=0}^{k-1} ( s + 1 - \frac{1}{2} k ) \Pi_{k-s}
\eqno (27)
$$
for a fractional supersymmetric oscillator. The energy spectrum of $H$ is
made of equally spaced levels with a ground state (singlet), a first
excited state (doublet), a second excited state (triplet), $\cdots$, 
a ($k-2$)-th excited state (($k-1$)-plet) followed by an infinite sequence of 
further excited states (all $k$-plets), see Section 4.3. 

(v) In the case where the algebra 
$W_k$ is restricted to $U_q(sl_2)$, see Appendix A, the corresponding
Hamiltonian $H$ is given by Eq.~(23) where the 
$f_{t}$ are given in Appendix A. This yields 
$$
 H = (k-1) J_+ J_-                                                       
   + \frac{1}{\sin {\frac{2 \pi  }{k}}} \sum_{s = 3}^{k  } 
       \sum_{t =   2}^{s-1} (t - 1) \> \sin {\frac{4 \pi t}{k}}  \Pi_{s}   
$$
$$   
   + \frac{1}{\sin {\frac{2 \pi  }{k}}} \sum_{s =   1}^{k-1} 
       \sum_{t =   s}^{k-1} (t - k) \> \sin {\frac{4 \pi t}{k}}  \Pi_{s}.
       \eqno (28)
$$ 
Alternatively, Eq.~(28) can be rewritten 
in the form (26) where $X_{\pm} \equiv J_{\pm}$ and $N \equiv  J_3$ 
and where the function $G$ is defined by
$$
G(X) = - [2X]_q,
$$
where the symbol $[ \ ]_q$ is defined by 
$$
[2X]_q = \frac{ q^{2X}  -  q^{-2X} }{ q - q^{-1} }
% \eqno (22bis(lamettre))
$$
with $X$ an arbitrary operator or number. The 
quadratic term $J_+J_-$ can be expressed in term of the Casimir operator $J^2$
of $U_q(sl_2)$, see Appendix A. Thus,
the so-obtained expression for the 
Hamiltonian $H$ is a simple function of $J^2$ and $J_3$. 

\subsection{A connection between fractional sQM and ordinary sQM}
In order to establish a connection between {\em fractional}
sQM of order $k$ and {\em ordinary} sQM (corresponding to $k = 2$),
it is necessary to construct subsystems from the doublet $(H, Q)_k$ 
that correspond to ordinary supersymmetric quantum-mechanical systems.
This may be achieved in the following way [50]. Equation (23) can be 
rewritten as 
$$
H = \sum_{s=1}^{k} H_{s} \> \Pi_{s}
\eqno (29) 
$$
where 
$$
H_s \equiv H_s(N) 
    = (k-1) F(N) - \sum_{t=2}^{k-1} (t-1) \> f_t(N-s+t)
    + (k-1)        \sum_{t=s}^{k-1}          f_t(N-s+t). 
\eqno (30)
$$
It can be shown that the operators 
$H_k \equiv H_0, H_{k-1}, \cdots, H_1$, turn out to be isospectral operators. 
By introducing 
$$
X(s)_- = 
% \sum_{n=1}^{d  } [H_s(n)  ]^{\frac{1}{2}} | n-1 , s-1 \rangle \langle n , s   |
  \sum_{n  }       [H_s(n)  ]^{\frac{1}{2}} | n-1 , s-1 \rangle \langle n , s   |
$$
$$
X(s)_+ = 
% \sum_{n=0}^{d-1} [H_s(n+1)]^{\frac{1}{2}} | n+1 , s   \rangle \langle n , s-1 |
  \sum_{n  }       [H_s(n+1)]^{\frac{1}{2}} | n+1 , s   \rangle \langle n , s-1 |
$$
it is possible to factorize $H_s$ as 
$$
H_s = X(s)_+ \> X(s)_-
% \eqno (17)
$$
modulo the omission of the ground state $|0,s\rangle$ 
% (which amounts to take the corresponding eigenvalue  
% as the zero energy for the spectrum of $H_s$). 
  (which amounts to substract the corresponding eigenvalue 
  from the spectrum of $H_s$).
Let us now define: (i) the two (supercharge) operators
$$
q(s)_- = X(s)_- \> \Pi_s, \quad 
q(s)_+ = X(s)_+ \> \Pi_{s-1}
% \eqno (18)
$$ 
and (ii) the (Hamiltonian) operator 
$$
h(s) = X(s)_- \> X(s)_+ \> \Pi_{s-1}   +   X(s)_+ \> X(s)_- \> \Pi_s.
\eqno (31)
$$
It is then a simple matter of calculation to prove that 
$h(s)$ is self-adjoint and that
$$
q(s)_+ = q(s)_-^{\dagger},             \quad
q(s)_{\pm}^2 = 0,                      \quad 
h(s) = \{ q(s)_- , q(s)_+ \},          \quad 
[ h(s) , q(s)_{\pm} ] = 0. 
% \eqno (20)
$$
Consequently, the doublet $(h(s), q(s))_2$, with
$q(s) \equiv q(s)_-$, 
satisfies Eq.~(18) with $k=2$ and thus 
defines an ordinary sypersymmetric 
quantum-mechanical system (corresponding to $k=2$).  

The Hamiltonian $h(s)$ is amenable to a form more 
appropriate for discussing the link between ordinary 
sQM and fractional sQM. Indeed, we can show that 
$$
% H_{s-1} = X(s)_- \> X(s)_+ 
            X(s)_- \> X(s)_+ = H_s(N+1).
\eqno (32)
$$
Then, by combining Eqs.~(2), (30) and (32),  
% by introducing Eqs.~(2) and (3) into Eq.~(15), 
Eq.~(31) leads to the important relation 
$$
h(s) = H_{s-1} \> \Pi_{s-1} + H_{s} \> \Pi_{s}
\eqno (33)
$$
to be compared with the expansion of $H$ in terms 
of supersymmetric partners $H_s$ (see Eq.~(29)). 

As a result, the system $(H,Q)_k$, corresponding to $k$-fractional
sQM, can be described in terms of $k-1$ sub-systems $(h(s),q(s))_2$, 
corresponding to ordinary sQM. The Hamiltonian $h(s)$ is given 
as a sum involving the supersymmetric partners 
$H_{s-1}$ and $H_s$ (see Eq.~(33)). Since the 
supercharges $q(s)_{\pm}$ commute with the Hamiltonian $h(s)$, 
it follows that 
$$
H_{s-1} X(s)_- = X(s)_- H_{s  }, \quad 
H_{s  } X(s)_+ = X(s)_+ H_{s-1}.
$$
As a consequence, the operators 
          $X(s)_+$ 
      and $X(s)_-$ render 
possible to pass from the spectrum of $H_{s-1}$ and $H_{s  }$
                        to the one of $H_{s}  $ and $H_{s-1}$, respectively. This 
result is quite familiar for ordinary sQM (corresponding to $s=2$).

For $k=2$, the operator $h(1)$ is nothing but the total Hamiltonian $H$ 
corresponding to ordinary sQM. For
arbitrary $k$, the other operators $h(s)$ are simple replicas (except for the
ground state of $h(s)$) of $h(1)$. In 
this sense, fractional sQM of order $k$
can be considered as a set of $ k-1 $ replicas of ordinary sQM 
corresponding to $k=2$ and 
typically described by $(h(s),q(s))_2$. As a further argument, 
it is to be emphasized that 
$$
H = q(2)_- \>
    q(2)_+ + \sum_{s=2}^{k} 
    q(s)_+ \> 
    q(s)_-
$$
which can be identified to $h(2)$ for $k=2$.

\section{A fractional supersymmetric oscillator}

\subsection{A special case of $W_k$}
In this section, we deal with the particular 
case where $f_s = 1$ and the deformed 
bosons $b(s)_{\pm} \equiv b_{\pm}$ are independent of $s$ 
with $s = 0, 1, \cdots, k-1$. We thus end 
up with a pair ($b_- , b_+$) of ordinary bosons, satisfying $[b_-,b_+] = 1$, 
    and a pair ($f_- , f_+$) of 
$k$-fermions. The ordinary bosons $b_{\pm}$ and the $k$-fermions $f_{\pm}$ 
may be considered as originating from the decomposition of a pair of 
$Q$-uons when $Q$ goes to the root of unity $q$ (see Appendix B).

Here, the two operators $X_-$ and $X_+$ are 
given by Eqs.~(9) and (10), where now
$b_{\pm}$ are ordinary boson operators. They satisfy the commutation relation
$[X_- , X_+] = 1$. Then, the number operator $N$ may defined by
$$
N = X_+ X_-, 
\eqno (34{\rm a})
$$
which is amenable to the form
$$
N  = b_+ b_-.
\eqno (34{\rm b})
$$
Finally, the  grading  operator  $K$  is still defined by Eq.~(11).
We can check that the operators $X_-$, $X_+$, $N$ and $K$ 
so-defined generate the generalized
Weyl-Heisenberg algebra $W_k$ defined by  Eq.~(2)  with $f_s = 1$ for
$s = 0, 1, \cdots, k-1$. The latter algebra $W_k$ can thus be realized 
with multilinear forms involving ordinary boson operators 
$b_{\pm}$ and $k$-fermion operators $f_{\pm}$. 

\subsection{The resulting fractional supersymmetric oscillator}
\noindent The supercharge operators $Q_-$ 
and $Q_+$ as well as the Hamiltonian $H$ 
associated with the algebra $W_k$ 
% introduced in Section~4.2???
can be constructed, 
in terms of the operators $b_-$, $b_+$, $f_-$ and $f_+$, 
according to the prescriptions given in Section~3.3. This
leads to the expression
$$
H = (k-1) b_+ b_-  +  (k-1) \sum_{s=0}^{k-1} ( s + 1 - \frac{1}{2} k ) \Pi_{k-s}
$$
to be compared with Eq.~(27). 

Most of the 
properties of the Hamiltonian $H$ are essentially
the same as the ones given above for the Hamiltonian (27). 
In particular, we can write (see Eqs.~(29) and (30))
$$
H = \sum_{s = 1}^{k} H_s \Pi_s, \quad 
                     H_s = (k-1) \left( b_+ b_- + \frac{1}{2} k + 1 - s \right)
$$
and  thus $H$  is  a linear combination of
projection operators with coefficients $H_s$ corresponding to isospectral Hamiltonians
(remember that $\Pi_k = \Pi_0$). 

To close this section, 
let us mention that the fractional supercoherent state 
$| z , \theta )$ defined in Appendix B is a coherent state 
corresponding to the Hamiltonian $H$. As a point of fact, we  
can check that the action of the evolution operator
$\exp (- {\rm i} H t)$ on the state $| z , \theta )$ gives
$$
\exp (- {\rm i} H t) \> | z , \theta ) =
\exp \left[ - \frac{{\rm i}}{2} (k-1) (k+2) t \right] 
                     \> | {\rm e}^{-{\rm i} (k-1) t} z , 
                          {\rm e}^{+{\rm i} (k-1) t} \theta ),
$$
i.e., another fractional supercoherent state. 

\subsection{Examples}
\subsubsection{Example 1}
As a first example, we take $k=2$, i.e., $q=-1$. Then, the
operators
$$
X_{\pm} = b_{\pm} \left( f_-   +    f_+ \right) 
$$
\noindent and the operators $K$ and $N$, see Eqs.~(11) and (34), are defined
in terms of bilinear forms of ordinary bosons 
$(b_- , b_+)$ and             ordinary fermions
$(f_- , f_+)$. The operators $X_-$, $X_+$, $N$ and $K$ satisfy 
$$
 [X_- , X_+     ]   = 1, \quad 
 [N   , X_{\pm} ]   = {\pm} X_{\pm}, 
$$
$$
 [K   , X_{\pm} ]_+ = 0, \quad
 [K   , N       ]   = 0, \quad K^2 = 1,
$$
\noindent which reflect bosonic  and  fermionic 
degrees of freedom, the bosonic degree corresponding to the triplet 
($X_- ,  X_+ , N$) and the fermionic degree to the Klein involution 
operator $K$. The projection operators 
$$
  \Pi_0 = \frac{1}{2}(1+K) = 1 - f_+ f_-, \quad
  \Pi_1 = \frac{1}{2}(1-K) =     f_+ f_-
$$
\noindent are here simple chirality operators and the supercharges 
$$
  Q_- = X_- \Pi_0 = b_- f_+, \quad
  Q_+ = X_+ \Pi_1 = b_+ f_-
$$
\noindent have the property
$$
Q_- ^2 = 
Q_+ ^2 = 0. 
$$
\noindent The Hamiltonian $H$ assumes the form
$$
H = \{ Q_- , Q_+ \} 
$$
\noindent which can be rewritten as
$$
H = b_+ b_- \Pi_0   +   b_- b_+ \Pi_1. 
$$
\noindent It is clear that the operator 
$H$ is self-adjoint and commutes with $ Q_- $ 
and   $Q_+$. Note that we recover that 
$Q_-$, $Q_+$ and $H$ span the Lie superalgebra 
$s \ell (1/1)$. We have
$$
H = b_+ b_-    +   f_+ f_-
$$
\noindent so that $H$ acts on the $Z_2$-graded space 
${\cal F} = {\cal F}_0 \oplus {\cal F}_1$. 
The operator $H$ corresponds to the ordinary or $Z_2$-graded 
supersymmetric oscillator whose energy spectrum $E$ is (in a symbolic way)
$$
E = 1 \oplus 2 \oplus 2 \oplus \cdots
$$ 
\noindent with equally spaced levels, the ground state  being  a 
singlet (denoted by 1) and all the excited states (viz., an infinite sequence) 
being doublets
(denoted by 2). Finally, note that the fractional supercoherent state
$ | z , \theta ) $ of Appendix B with $k=2$ is a coherent state for the
Hamiltonian $H$ (see also Ref.~[53]). 
 
\subsubsection{Example 2}
We continue with $k=3$, i.e., 
$$
q = \exp \left( \frac{2 \pi {\rm i}}{3} \right).
$$
In this case, we have 
$$
X_- 
= b_- \left( f_-   +    \frac{ f_+ ^{2}}{[[2]]_q !} \right)
= b_- \left( f_- - q f_+^2 \right),
$$
$$
X_+ 
= b_+ \left( f_-   +    \frac{ f_+ ^{2}}{[[2]]_q !} \right)^{2}
= b_+ \left( f_+ + f_-^2 + q^2 f_+^2 f_- \right).
$$
\noindent Furthermore, $K$ and $N$ are given by (11) and (34), 
where here $( b_- , b_+ )$ are ordinary bosons 
and $( f_- , f_+ )$ are  $3$-fermions. We hence have 
$$
 [ X_- , X_+ ] = 1, \quad
 [ N , X_{\pm} ] = {\pm} X_{\pm},
$$
$$
 [ K , X_+ ]_q  = [ K , X_- ]_{\bar q} = 0, \quad
 [K , N] = 0,         \quad  K^3 = 1.
$$
\noindent Our general definitions can be specialized to
\begin{eqnarray*}
\Pi_0 &=& \frac{1}{3} \left( 1 + q^3 K + q^3 K^2 \right) \\
\Pi_1 &=& \frac{1}{3} \left( 1 + q^1 K + q^2 K^2 \right) \\
\Pi_2 &=& \frac{1}{3} \left( 1 + q^2 K + q^1 K^2 \right)          
\end{eqnarray*}
or equivalently
\begin{eqnarray*}
\Pi_0 &=& 1 + (q-1) f_+ f_-  -  q    f_+ f_- f_+ f_- \\
\Pi_1 &=&       - q f_+ f_-  + (1+q) f_+ f_- f_+ f_- \\
\Pi_2 &=&           f_+ f_-  -       f_+ f_- f_+ f_-
\end{eqnarray*}
\noindent for the projection operators and to 
$$
Q_- = X_- (\Pi_0 + \Pi_2) = b_- f_+ \left( f_-^2 -  q f_+   \right) 
$$
$$ 
Q_+ = X_+ (\Pi_1 + \Pi_2) = b_+     \left( f_-   -  q f_+^2 \right) f_- 
$$ 
\noindent for the supercharges with the property
$$
Q_- ^3 = 
Q_+ ^3 = 0. 
$$
\noindent By introducing the Hamiltonian $H$ via
$$
Q_- ^{2} Q_+     +   Q_- Q_+ Q_- 
                               +   Q_+ Q_- ^{2}
                               =   Q_- H
$$
\noindent we obtain 
$$
H = \left( 2 b_+ b_-  -  1 \right) \Pi_0 +
    \left( 2 b_+ b_-  +  3 \right) \Pi_1 +
    \left( 2 b_+ b_-  +  1 \right) \Pi_2
$$
\noindent which acts on the $Z_3$-graded space 
${\cal F} = {\cal F}_0 \oplus {\cal F}_1 \oplus {\cal F}_2$ 
and can be rewritten as 
$$
H = 2 b_+ b_-  -  1  +  2(1 - 2q) f_+ f_-  +  2(1 + 2 q) f_+ f_- f_+ f_-
$$
\noindent in terms of boson and 3-fermion operators.
We can check that the operator $H$ is self-adjoint and 
commutes with $ Q_- $ 
and   $Q_+$. The energy spectrum of $H$ reads 
$$
E = 1 \oplus 2 \oplus 3 \oplus 3 \oplus \cdots.
$$ 
\noindent It contains equally spaced levels with a 
nondegenerate ground state (denoted as 1), a doubly 
degenerate first excited state (denoted as 2) and an infinite
sequence of triply degenerate excited states (denoted as 3). 

\section{Differential realizations}
In this section, we consider the case of the algebra $W_k$ defined by
Eqs.~(2b)-(2e) and Eq.~(14) with $c_0 = 1$ and $c_s = c \delta (s,1)$, 
$c \in {\bf R}$, for $s = 1, 2, \cdots, k-1$. In other words, we have
$$
 [X_- , X_+] = 1 + c K, \quad K^k = 1,
\eqno (35{\rm a})
$$
$$
 [K , X_+]_q = [K , X_-]_{\bar q} = 0,
\eqno (35{\rm b}) 
$$
which corresponds to the $C_{\lambda}$-extended oscillator. The operators
$X_-$, $X_+$ and $K$ can be realized in terms of a bosonic variable $x$ and
its derivative $ \frac{d}{dx} $ satisfying
$$
[ \frac{d}{dx}  , x] = 1
$$
and a $k$-fermionic variable (or generalized 
Grassmann variable) $\theta$ and its
derivative $ \frac{d}{d \theta} $ satisfying [24,36] 
(see also Refs.~[25-31])
$$
[ \frac{d}{d \theta}  , \theta]_{\bar q} = 1, \quad 
{\theta}^k = \left( { \frac{d}{d \theta} } \right)^k=0.
$$
Of course, the sets $\{ x ,  \frac{d}{dx}  \}$ and $\{ \theta ,  \frac{d}{d \theta}  \}$
commute. It is a simple matter of calculation to derive the two following
identities
$$
\left(  \frac{d}{d \theta}  + \frac{\theta^{k-1}}{[[k-1]]_{\bar q}!} \right)^k = 1
$$
and
$$
(  \frac{d}{d \theta}  \theta - \theta  \frac{d}{d \theta}  )^k = 1,
$$
which are essential for the realizations given below. 

As a first realization, we can show that the shift operators
$$
X_- =  \frac{d}{dx}  \left(  \frac{d}{d \theta}  + \frac{\theta^{k-1}}{[[k-1]]_{\bar q}!} \right)^{k-1} 
    - \frac{c}{x} \theta,
$$
$$
X_+ = x \left(  \frac{d}{d \theta}  + \frac{\theta^{k-1}}{[[k-1]]_{\bar q}!} \right),
$$
and the Witten grading operator
$$
K = [  \frac{d}{d \theta}  , \theta ] 
$$
satisfy Eq.~(35). This realization of $X_-$, $X_+$ and $K$ 
clearly exibits the bosonic and $k$-fermionic degrees of 
freedom via the sets $\{ x ,  \frac{d}{dx}  \}$ and $\{ \theta ,  \frac{d}{d \theta}  \}$,
respectively. In the particular case $k=2$, the $k$-fermionic 
variable $\theta$ is an ordinary Grassmann variable and the 
supercharge operators $Q_-$ and $Q_+$ take the simple form
$$
Q_- = \left(  \frac{d}{dx}  - \frac{c}{x} \right) \theta, 
\eqno (36{\rm a})
$$
$$
Q_+ = x  \frac{d}{d \theta}.
\eqno (36{\rm b})
$$
(Note that the latter realization for $Q_-$ and $Q_+$ is
valid for $k = 3$ too.)

Another possible realization of $X_-$ and $X_+$ 
for arbitrary $k$ is 
$$
X_- = P \left(  \frac{d}{d \theta}  + \frac{\theta^{k-1}}{[[k-1]]_{\bar q}!} \right)^{k-1} 
    - \frac{c}{x} \theta,
$$
$$
X_+ = X \left(  \frac{d}{d \theta}  + \frac{\theta^{k-1}}{[[k-1]]_{\bar q}!} \right),
$$
where $P$ and $X$ are the two canonically conjugated quantities
$$
P = \frac{1}{\sqrt{2}} \left(  x + \frac{d}{dx} - \frac{c}{2x} K \right)
$$
and
$$
X = \frac{1}{\sqrt{2}} \left(  x - \frac{d}{dx} + \frac{c}{2x} K \right).
$$
This realization is more convenient for a Schr\"odinger type approach
to the supersymmetric Hamiltonian 
$H$. According to Eq.~(23), 
we can derive an Hamiltonian 
$H$ involving bosonic and $k$-fermionic degrees of freedom.
To illustrate this point, let us 
continue with the particular case $k=2$. It can be seen that 
the supercharge operators (36) must be replaced by 
$$
Q_- = \left( P - \frac{c}{X} \right) \theta,
$$
$$
Q_+ = X  \frac{d}{d \theta}.
$$
(Note the formal character of $Q_-$ since the 
definition of $Q_-$ lies on the existence 
of an inverse for the operator $X$.) Then, 
we obtain the following Hamiltonian 
$$
H = - \frac{1}{2} \left[ \left( \frac{d}{dx} - \frac{c}{2 x} K \right)^2 
    - x^2 + K + c (1 + K) \right].
$$
For $c=0$, we have (cf.~Ref.~[8])
$$
H = - \frac{1}{2} \frac{d^2}{dx^2} 
    + \frac{1}{2} x^2  
    - \frac{1}{2} K 
$$
that is the Hamiltonian for an ordinary super-oscillator, i.e., a 
$Z_2$-graded supersymmetric oscillator. Here, the bosonic character
arises from the bosonic variable $x$ and the fermionic    character
       from the ordinary Grassmann variable $\theta$ in $K$. 

\section{Closing remarks} 
The basic ingredient for the present work is the definition of 
a generalized Weyl-Heisenberg algebra $W_k$ that depends on $k$
structure constants $f_s$ ($s = 0, 1, \cdots, k-1$). We have 
shown how to construct ${\cal N}=2$ fractional supersymmetric 
Quantum Mechanics of order $k$, $k \in \{ 0 , 1 \}$, by means 
of this $Z_k$-graded algebra $W_k$. The ${\cal N}=2$ 
dependent supercharges and a general Hamiltonian are derived in 
terms of the generators of $W_k$. This general fractional supersymmetric 
Hamiltonian is a linear combination of isospectral supersymmetric 
partners $H_0, H_{k-1}, \cdots, H_{1}$ and this result is at the root 
of the development of fractional supersymmetric Quantum Mechanics 
of order $k$ as a set of replicas of ordinary supersymmetric Quantum 
Mechanics (corresponding to $k=2$). The general Hamiltonian covers 
various dynamical systems corresponding to translational and cyclic 
shape-invariant potentials. A special attention has been given
to the fractional supersymmetric oscillator. From a general point of view, 
the formalism presented in this paper is useful for studying 
exact integrable quantum systems and for constructing their 
coherent states.

\section*{Acknowledgments}
This paper is dedicated to the memory of 
the late Professor Per-Olov L\"owdin. The 
senior author (M.R.~K.) has had a chance to 
benefit from conversations with Professor L\"owdin. His
lectures and papers are a model for many people from the Quantum Physics 
and Quantum Chemistry communities. He will remain an example for many of us. 

One of the authors (M.~D.) would like to thank 
the Institut de Physique Nucl\'eaire de Lyon 
for the kind hospitality extended to him 
at various stages (during 1999-2003) 
of the development of this work.

\section*{Appendix A: Connection between $W_k$ and $U_q(sl_2)$}
Let us now show that the quantum algebra $U_q(sl_2)$, 
with $q$ being the $k$-th root of unity
given by (1), turns out to be a particular form of $W_k$. The algebra
$U_q(sl_2)$ is spanned by the generators
$J_-$,  $J_+$,  $q^{J_3}$  and  $q^{-J_3}$
that satisfy the relationships 
$$
[J_+ , J_-] =  [2 J_3]_q,
$$ 
$$
q^{J_3} J_+ q^{-J_3} =       q  J_+, \quad
q^{J_3} J_- q^{-J_3} = {\bar q} J_-,
$$
$$
q^{J_3} q^{-J_3} = q^{-J_3} q^{J_3} = 1.  
$$
% where the symbol $[ \ ]_q$ is defined by (22bis???laisser).
It is straightforward to prove that the operator
$$
J^2 = J_- J_+  +  \frac{q^{+1} q^{2J_3} + q^{-1} q^{-2J_3}} {(q - q^{-1})^2}
$$ 
or
$$
J^2 = J_+ J_-  +  \frac{q^{-1} q^{2J_3} + q^{+1} q^{-2J_3}} {(q - q^{-1})^2}
$$ 
is an invariant of $U_q(sl_2)$. In view of Eq.~(1), the operators $J_-^k$,
$J_+^k$, 
$( q^{ J_3} )^k$, and 
$( q^{-J_3} )^k$ belong, likewise $J^2$, to the center of $U_q(sl_2)$. 

In the case where the deformation parameter $q$ is 
a root of unity,
the representation theory of $U_q(sl_2)$ is richer than
the one for $q$ generic. The algebra $U_q(sl_2)$ admits 
finite-dimensional 
representations of dimension $k$ such that
$$
J_-^k = A, \quad 
J_+^k = B,
$$
where $A$ and $B$ are constant matrices. Three types of representations
have been studied in the literature [49]:

(i)   $A = B = 0$ (nilpotent representations),

(ii)  $A = B = 1$ (cyclic or periodic representations),

(iii) $A = 0$ and $B = 1$ or 
      $A = 1$ and $B = 0$
      (semi-periodic representations).
      
\noindent Indeed, the realization of fractional sQM
based on $U_q(sl_2)$  does not depend of the choice (i), (ii) or (iii) 
in contrast with the work in 
Ref.~[36] where nilpotent representations 
corresponding to the choice (i) 
were considered. The only
important ingredient is to take 
$$
\left( q^{J_3} \right)^k = 1
$$
that ensures a $Z_k$-grading of the Hilbertean 
representation space of $U_q(sl_2)$. 

The contact with the algebra $W_k$ is established by putting
$$
X_{\pm} = J_{\pm}, \quad
N   = J_3,         \quad
K   = q^{J_3},
$$
and by using the definition (3) of $\Pi_s$ as function of $K$. 
Here, the operator $\Pi_s$ is a projection operator on the subspace,
of the representation space of $U_q(sl_2)$,
corresponding to a given eigenvalue of $J_3$. 
It is easy to check that the
operators $X_-$, $X_+$, $N$ and $K$ satisfy Eq.~(2) with
$$
f_s(N) = - [2s]_q = - \frac{\sin \frac{4 \pi s}{k}}{\sin \frac{2 \pi}{k}}
$$
for $s = 0, 1, \cdots, k-1$. The quantum algebra 
$U_q(sl_2)$, with $q$ given by (1), then appears as a further particular 
case of the generalized Weyl-Heisenberg algebra 
$W_k$.

\section*{Appendix B: A $Q$-uon $\to$ boson $+$ $k$-fermion decomposition}
We shall limit ourselves to give an outline of this decomposition (see 
      Dunne {\em et al.} [54]
and Mansour {\em et al.} [55] for an alternative and more
rigorous mathematical presentation based on the isomorphism between the
braided $Z$-line and the $(z , \theta)$-superspace). We start from a 
$Q$-uon algebra spanned by three operators $a_-$, $a_+$ and $N_a$ 
satisfying the relationships [18] 
(see also Refs.~[19-22])
$$
[a_- , a_+]_Q = 1, \quad
[ N_a , a_{\pm} ] = {\pm} a_{\pm},
$$
where $Q$ is generic (a real number different from zero). The
action of the operators $a_-$, $a_+ = a_-^{\dagger}$ 
and $N_a = N_a^{\dagger}$ on a Fock space
${\cal F} = \{ | n \rangle : n \in {\bf N} \}$ is given by 
$$
N_a | n \rangle = n | n \rangle,
$$
and
$$
a_- | n \rangle = ([[n + \sigma - \frac{1}{2}]]_Q)^{\alpha} \> | n - 1 \rangle,
$$
$$
a_+ | n \rangle = ([[n + \sigma + \frac{1}{2}]]_Q)^{\beta}  \> | n + 1 \rangle,
$$ 
where  
$\alpha + \beta = 1$ with $0 \leq \alpha \leq 1$ and 
$0 \leq \beta \leq 1$. For $\alpha = \beta = \frac{1}{2}$, 
let us consider the $Q$-deformed Glauber coherent 
state [18] (see also Ref.~[56])
$$
| Z ) = \sum_{n=0}^{\infty} 
\frac{ \left( Z a_+ \right)^n }{ [[n]]_Q! } \> | 0 \rangle
       = \sum_{n=0}^{\infty} 
   \frac{ Z^n }{ ([[n]]_Q!)^{\frac{1}{2}} } \> | n \rangle
$$ 
\noindent (with $Z \in {\bf C}$). If we do the replacement 
$$
Q \leadsto q = \exp \left( \frac{2 \pi {\rm i}}{k} \right), \quad 
k \in {\bf N} \setminus \{ 0,1 \},
$$
then we have 
$\lbrack\lbrack k \rbrack\rbrack_Q ! \to 
 \lbrack\lbrack k \rbrack\rbrack_q ! = 0$. Therefore,
in order to give a sense to $| Z )$ for $Q \leadsto q$, we have to do the replacement
$$
a_+ \leadsto f_+ \quad {\rm with} \quad f_+^k = 0,
$$
$$
a_- \leadsto f_- \quad {\rm with} \quad f_-^k = 0.
$$
\noindent We 
thus end up with what we call a $k$-fermionic algebra $F_k$ spanned 
by the operators  $f_-$, $f_+$ and $N_f \equiv N_a$ completed by the
adjoints
$f_+^{\dagger}$ and 
$f_-^{\dagger}$ of $f_+$ and $f_-$, 
respectively [31,33]. The defining
relations for the $k$-fermionic algebra $F_k$ are
$$
 [ f_- , f_+ ]_q = 1, \quad
 [ N_f , f_{\pm} ] = {\pm} f_{\pm}, \quad
 f_-^k = 
 f_+^k = 0,
$$
$$
[ f_+^{ \dagger} , 
  f_-^{ \dagger} ]_{\bar q} = 1, \quad
 [ N_f , f_{\pm}^{ \dagger} ] = 
   {\mp} f_{\pm}^{ \dagger}, \quad
 \left( f_-^{ \dagger} \right)^k = 
 \left( f_+^{ \dagger} \right)^k = 0,
$$
$$
 f_- f_+^{ \dagger} - q^{- \frac{1}{2}} f_+^{ \dagger} f_- = 0, \quad
 f_+ f_-^{ \dagger} - q^{+ \frac{1}{2}} f_-^{ \dagger} f_+ = 0.
$$
The case $k=2$ corresponds 
to ordinary fermion operators and the case $k \to \infty$ 
to ordinary boson   operators. 
In the two 
latter cases,  we can take $f_- \equiv f_+^{\dagger}$ and 
                           $f_+ \equiv f_-^{\dagger}$; in the 
other cases, the consideration of the two couples 
$(f_-, f_+^{\dagger})$ and 
$(f_+, f_-^{\dagger})$ is absolutely necessary. In the case 
where $k$ is arbitrary, we shall speak of $k$-fermions. 
The $k$-fermions are objects interpolating
between
fermions and bosons. They share some properties with the 
para-fermions [24,25,27]
and the anyons as introduced by 
Goldin {\em et al.} [10]
(see also Ref.~[9]). If we define 
$$
b_{\pm} = \lim_{Q \leadsto q} \frac{    a_{\pm}^k    }
                               { \left( [[k]]_Q ! \right)^{\frac{1}{2}}} 
$$
we obtain 
 $$
 [ b_- , b_+ ] = 1
 $$
so that the operators $b_-$ and $b_+$ can be considered as ordinary boson
operators. 
This is at the root of the two following results [31].

As a first result, the set $\{ a_- , a_+ \}$ gives rise, for $Q \leadsto q$, 
to two commuting sets: 
The set $\{ b_- , b_+ \}$ of boson operators and the set of $k$-fermion
operators $\{ f_- , f_+ \}$. As a second result, this decomposition leads 
to the replacement of the $Q$-deformed coherent state $ | Z ) $ by the 
so-called fractional supercoherent state
$$
| z , {\theta} ) = \sum_{n=0}^{\infty} 
                    \sum_{s=0}^{k-1} 
                    \frac{ \theta^s }{ ([[s]]_q!)^{\frac{1}{2}} }
                    \> \frac{ z^n }{ \sqrt{n!} }
                    \> | n , s \rangle,
$$
\noindent where $z$ is a (bosonic) complex variable and $\theta$ a 
($k$-fermionic) generalized Grassmann 
variable [24,27,36,57] 
with $\theta^k = 0$. The fractional 
supercoherent state $| z , {\theta} )$ 
is an eigenvector of the product $f_- b_-$ with the eigenvalue 
$z \theta$. The state $| z^k , {\theta} )$ can be seen to 
be a linear combination of the coherent states introduced by 
Vourdas [58] 
with coefficients in the generalized Grassmann algebra spanned by $\theta$ and
the derivative $ \frac{d}{d \theta} $. 
   
In the case $k=2$, the state $| z , {\theta} )$
turns out to be a coherent state for the ordinary (or $Z_2$-graded) 
supersymmetric oscillator [53]. For $k \ge 3$, the state $| z , {\theta} )$
is a coherent state for the $Z_k$-graded supersymmetric oscillator (see
Section~4). 

It is possible to find a realization of the operators $f_-$, 
                                                      $f_+$, 
                                                      $f_+^{\dagger}$ and
                                                      $f_-^{\dagger}$ in terms of          
Grassmann variables $(\theta, {\bar \theta})$ and 
their $q$- and ${\bar q}$-derivatives
$(\partial_{\theta}, \partial_{\bar \theta})$. We take           
Grassmann variables 
$\theta$ and ${\bar \theta}$ such that $ \theta^k = {\bar \theta}^k = 0 $
[24,27,36,57]. The sets
$\{ 1,       \theta , \cdots,       \theta  ^{k-1} \}$ and 
$\{ 1, {\bar \theta}, \cdots, {\bar \theta} ^{k-1} \}$ 
span the  same  Grassmann algebra $\Sigma_k$.  
The $q$- and ${\bar q}$-derivatives are 
formally defined by 
\begin{eqnarray*}
\partial_\theta f(\theta) &=& {f(q\theta) - f(\theta) \over (q - 1) \theta}, \\
\partial_{\bar \theta} g({\bar \theta}) &=& 
{g({\bar q} {\bar \theta}) - g({\bar \theta}) \over ({\bar q} - 1) {\bar \theta}}.
\end{eqnarray*}
\noindent Therefore, by taking 
% \begin{eqnarray*}
$$
f_+   =                        \theta ,    \   
f_-   = \partial_              \theta ,    \ 
f_-^{\dagger} =          {\bar \theta},    \   
f_+^{\dagger} = \partial_{\bar \theta},
$$
% \end{eqnarray*}
\noindent we have
% \begin{eqnarray*}
$$
\partial_{\theta} \theta - q \theta \partial_{\theta} = 1,           \ 
\left( \partial_{\theta} \right)^k   =  \theta ^k     = 0,
$$
$$
\partial_{\bar \theta} {\bar \theta} - {\bar q} {\bar \theta} \partial_{\bar \theta} = 1,         \ 
\left( \partial_{\bar \theta} \right)^k   =   {\bar \theta} ^k                       = 0,
$$
$$ 
\partial_{\theta} \partial_{\bar \theta} - q^{-{1 \over 2}} \partial_{\bar \theta} 
\partial_{\theta} = 0,  \ 
\theta {\bar \theta}                     - q^{+{1 \over 2}} {\bar \theta} \theta      
                  = 0.
$$
% \end{eqnarray*}
\noindent Following Majid and Rodr\'\i guez-Plaza [57], we define the integration 
process
$$
\int d{      \theta}  \> {      \theta} ^n = 
\int d{\bar {\theta}} \> {\bar {\theta}}^n = 0  \  {\hbox{for}}  \  n = 0, 1, \cdots, k-2
$$
and
$$
\int d{\theta} \> {\theta}^{k-1} = \int d{\bar {\theta}} \> {\bar {\theta}}^{k-1} = 1
$$
\noindent which gives the Berezin integration for the particular case $k=2$.

% \section*{References IJQC}

\end{document}